# Spin accumulation detection of FMR driven spin pumping in silicon-based metal-oxide-semiconductor heterostructures

Y. Pu[1], P. M. Odenthal[2], R. Adur[1], J. Beardsley[1], A. G. Swartz[2], D. V. Pelekhov[1], R. K. Kawakami[2], J. Pelz[1], P. C. Hammel[1], E. Johnston-Halperin[1]

[1]Department of Physics, The Ohio State University, Columbus, Ohio 43210

[2]Department of Physics and Astronomy, University of California, Riverside, California 92521

**The use of the spin Hall effect and its inverse to electrically detect and manipulate dynamic spin currents generated via ferromagnetic resonance (FMR) driven spin pumping has enabled the investigation of these dynamically injected currents across a wide variety of ferromagnetic materials. However, while this approach has proven to be an invaluable diagnostic for exploring the spin pumping process it requires strong spin-orbit coupling, thus substantially limiting the materials basis available for the detector/channel material (primarily Pt, W and Ta). Here, we report FMR driven spin pumping into a weak spin-orbit channel through the measurement of a spin accumulation voltage in a Si-based metal-oxide-semiconductor (MOS) heterostructure. This alternate experimental approach enables the investigation of dynamic spin pumping in a broad class of materials with weak spin-orbit coupling and long spin lifetime while providing additional information regarding the phase evolution of the injected spin ensemble via Hanle-based measurements of the effective spin lifetime.**

The creation and manipulation of non-equilibrium spin populations in non-magnetic materials (NM) is one of the cornerstones of modern spintronics. These excitations have to date relied primarily on charge based phenomena, either via direct electrical injection from a ferromagnet (FM)[1-7] or through the exploitation of the spin-orbit interaction[8-10]. Ferromagnetic resonance (FMR) driven spin pumping[11-22] is an emerging method to dynamically inject pure spin current into a NM with no need for an accompanying charge current, implying substantial potential impacts on low energy cost, high efficiency spintronics. However, while the creation of these non-equilibrium spin currents does not require a charge current, previous studies of transport-detected spin pumping do rely on a strong spin-orbit interaction in the NM to convert the spin current into a charge current in the detector via the inverse spin-Hall effect (ISHE) [12-20]. This approach has proven to be an instrumental diagnostic, but it does carry with it several limitations; specifically, the ISHE measures spin current not spin density, is only sensitive to a single component of the full spin vector and is only effective in materials with strong spin-orbit coupling. Here we demonstrate an alternate detection geometry relying on the measurement of a spin accumulation voltage using a ferromagnetic electrode, similar to the three-terminal geometry pioneered for electrically-driven spin injection[4,23-28]. This approach dramatically expands the materials basis for FMR driven spin pumping, allows for the direct measurement of spin accumulation in the channel and enables the phase-sensitive measurement of the injected spin population.

Our study is performed in a silicon-based metal-oxide-semiconductor (MOS) structure compatible with current semiconductor logic technologies. Using a Fe/MgO/p-Si tunnel diode we achieve spin pumping into a semiconductor across an insulating dielectric. This approach allows voltage-based detection of the spin accumulation under the electrode[4-6,23-29]. Further, we demonstrate sensitivity to the phase of the injected spin via the observation of Hanle dephasing in the presence of an out-of-plane magnetic field. These results establish a bridge between the pure spin currents generated by FMR driven spin pumping and traditional charge-based spin injection, laying the foundation for a new class of experimental probes and promising the development of novel spin-based devices compatible with current CMOS technologies.

Tunnel diodes are fabricated from Fe(10nm)/ MgO(1.3nm)/ Si(100) heterostructures grown by molecular beam epitaxy (MBE). The p-type Si substrates are semiconductor on insulator (SOI) wafers with a 3 μm thick Si device layer containing $5\times10^{18}$ cm$^{-3}$ boron dopants, producing a room temperature resistivity of $2\times10^{-2}$ Ωcm. The device is patterned by conventional photolithography techniques into a Fe/MgO/Si tunnel contact of 500 μm × 500 μm lateral size, placed 1 mm away from Au reference contacts for voltage measurements. Spin pumping and FMR measurements are performed in the center of a radio frequency (RF) microwave cavity with $f$ = 9.85GHz with a DC magnetic field, $H_{DC}$, applied along the $x$-axis, as sketched in Fig. 1a.  On resonance, a pure spin current is injected into the silicon channel via coupling between the precessing magnetization of the ferromagnet, $M$, and the conduction electrons in the silicon. This spin current induces an imbalance in the spin-resolved electrochemical potential and consequent spin accumulation given by $\Delta\mu_S = \mu_\uparrow - \mu_\downarrow$, where $\mu_\uparrow$ and $\mu_\downarrow$ are the chemical potentials of up and down spins, respectively. Using a standard electrical spin detection technique[4-6,23-29] the spin accumulation can be detected electrically via the relationship:

$$V_S = P\eta \frac{\Delta\mu_S}{2e} \qquad (1)$$

where $V_S$ is the spin-resolved voltage between Fe and Si, $P$ is the spin polarization of Fe, $\eta$ is the spin detection efficiency, and $\Delta\mu_S$ is assumed to be proportional to the component of the net spin polarization parallel to $\boldsymbol{M}$[30]. Figure 1c shows the magnetic field dependence of the FMR intensity (upper panel) and the spin accumulation induced voltage $V_S$ (lower panel), clearly demonstrating spin accumulation at the ferromagnetic resonance.

Figure 2a shows the RF power, $P_{RF}$, dependence of the FMR intensity (upper panel) and $V_S$ (lower panel) on resonance; the former is proportional to the square root of $P_{RF}$ and latter is linear with $P_{RF}$, consistent with ISHE detected spin pumping[18-20]. As shown in Fig.2b, $V_S$ is constant when $\boldsymbol{M}$ reverses, consistent with our local detection geometry wherein the injected spin is always parallel to the magnetization of the FM electrode. Note that this is in contrast to the magnetization dependence of ISHE detection, wherein the sign of the ISHE gives a measure of the spin orientation relative to the detection electrode. As a result, our technique distinguishes the spin accumulation signal from artifacts due to magneto-transport or spin transport, such as the anomalous Hall effect, ISHE or spin Seebeck effect, which depend on the direction of $\boldsymbol{M}$. In addition, the current-voltage characteristic (I-V) of the tunnel contact is linear at room temperature (see Supplementary Information), implying at best a weak rectification of any RF-induced pickup currents. This expectation is confirmed by the small offset (below 10μV) observed in spin accumulation voltage measurements. Our technique also rules out potential spurious signals due to magneto-electric transport such as tunneling anisotropic magneto-resistance (TAMR) that would require a rectified bias voltage of at least mV scale to give the observed μV scale signal observed at resonance.

In order to probe the dynamics of the observed spin accumulation, the applied magnetic field is rotated towards the sample normal (within the *xz*-plane) by an angle $\alpha$, introducing an out-of-plane component, $H_z$. Due to the strong demagnetization field of our thin-film geometry (~2.2 T) the orientation of the magnetization lags the orientation of the applied field, remaining almost entirely in-plane (the maximum estimated deviation is 2°). As a result, the injected spins (parallel to *M*) precess due to the applied field. As the magnitude of $H_z$ increases this precession will lead to a dephasing of the spin ensemble and consequent decrease in its net magnetization, (the Hanle effect[3-6,23-29]). Figure 3a shows the FMR spectrum for different angles $\alpha$; the resonant field $H_{FMR}$ changes from 250G to 400G as $\alpha$ changes from 0 to 40 degrees. This increase is consistent with the fact that the in-plane component of *H* primarily determines the resonance condition, so as $\alpha$ increases a larger total applied field is therefore required to drive FMR (see Supplementary Information). Figure 3b shows $V_S$ vs. $H_{DC}$ over the same angular range. The peak position of $V_S$ shifts in parallel with the FMR spectrum, but the peak value decreases with increasing $H_z$, as expected for Hanle-induced dephasing in an ensemble of injected spins[3-6,23-29].

For an isotropic ensemble of spins precessing in a uniform field perpendicular to *M* the effect of this dephasing on $V_S$ can be described by a simple Lorentzian function:

$$V_S \propto \Delta\mu_S \propto S_x(H_z) = \frac{S_0}{1+(\omega\tau)^2} \quad (2)$$

where $\omega$ is the Larmor frequency given by $\omega = g_{eff}\mu_B H_z/\hbar$, $g_{eff}$ is the effective Landé g-factor, $\mu_B$ is the Bohr magneton, $\hbar$ is the Plank's constant, and $\tau$ is the spin lifetime. In the more general case that *H* is not perpendicular to *M*, as is the case here, then Eq. (2) should be replaced by the more general function:

$$V_S \propto S_x = S_0\left\{\frac{\omega_x^2}{\omega_{total}^2} + \frac{\omega_y^2+\omega_z^2}{\omega_{total}^2}\frac{1}{1+(\omega_{total}\tau)^2}\right\} \quad (3)$$

where $\omega_i = H_i g_{eff}\mu_B/\hbar$ $(i=x,y,z)$ [23,27]. If *H* is in the *xz*-plane, this reduces to

$$V_S \propto S_x = S_0 \left\{ \cos^2 \alpha + \sin^2 \alpha \frac{1}{1+(\omega_{total}\tau)^2} \right\} \quad (4)$$

This general behavior has been observed in previous studies of three terminal electrical spin injection[4,23-28] (Fig. 4a). However, it has been widely reported in electrically detected spin injection experiments that spatially varying local fields due to the magnetic electrodes, coupling to interfacial spin states and other non-idealities generate spin dynamics that are not well described by this simple model. As a result $\tau$ is generally understood to represent an effective spin lifetime, $\tau_{eff}$, and while there are some initial efforts to more quantitatively account for the real sample environment, such as the so-called "inverted-Hanle" measurement[23,26,27], a detailed model of these interactions is currently lacking.

We explore the functional dependence of the dephasing of our FMR driven spin current by plotting the peak spin accumulation voltage, $V_S^{peak}$, as a function of $H_z$ (Fig. 4b, solid circles). The suppression of the spin accumulation at high magnetic fields seen in Fig. 3 is a clear indication of the dephasing of the injected spin ensemble; however, attempts to fit this behavior to Eq. (4) reveal that this simple, isotropic model fails to accurately reproduce our data (Fig. 4b, black dashed lines). In particular, it is a feature of the isotropic model that for a magnetic field that is not parallel to $z$ that the spin polarization along ***M***, and therefore the measured accumulation voltage, does not go to zero at high field even for infinite spin lifetime. This discrepancy likely arises from contributions due to the various non-idealities discussed above; in particular, as we discuss below, we believe that the coupling to localized states and the impact of bulk spin diffusion may play a more central role in this experimental geometry.

We note that our data is well described by a simple Lorentzian (Eq. 2), though the relationship between the effective spin lifetime extracted from this fit (0.6 ns), which we label $\tau_{FMR}$, to the $\tau_{eff}$ defined in Eq. (4) is not clear. For comparison, the intensity of the FMR signal is found to be constant to within roughly 10% (solid red triangles), suggesting that the spin current is roughly constant and indicating that FMR-driven heating[19,20], if present, does not contribute significantly to the field-induced suppression of $V_S^{peak}$ seen in Fig. 4b.

A key advantage of our experimental geometry is that it allows direct comparison of this dephasing with the more traditional three-terminal electrical injection *within the same device*[26]. Figure 4b (open purple circles) shows the spin accumulation voltage measured in the three-terminal geometry as a function of a perpendicular applied field, $H_z$. The dephasing in this geometry is clearly much slower than in FMR driven spin pumping. This observation is supported by the Lorentzian fit to Eq. (2) indicated by the solid purple line, yielding an effective lifetime of 0.11 ns, consistent with previous reports by our group and others[4,23-27]. In considering the origin of this discrepancy in observed lifetime a natural suspicion falls on the different experimental methodologies. Specifically, for the spin pumping case the field is applied at an angle $\alpha$, resulting in both in-plane and perpendicular components to the field, while for the DC current injection only the perpendicular component is present.

The consequence of this vector magnetic field is twofold: first, it will rotate the precession axis of the injected spins away from the perpendicular case implicit in the simple Hanle model as described above, and second, it will generate an "inverted" Hanle effect that has been proposed to derive from the interplay between an in-plane applied magnetic field and some finite inhomogeneity in the local fields due to the magnetization of the electrode. In Fig. 4c we explore this behavior in a control sample wherein we perform both traditional Hanle (i.e. wherein the only applied magnetic field is $H_z$) and rotating Hanle measurements (i.e. wherein the magnetic field is rotated by an angle $\alpha$, yielding both in-plane and perpendicular components to $H$), see Supplementary Information. The field values for the rotating Hanle experiment are chosen to correspond to the values of $H_z$ from Fig. 4b. As expected, the traditional Hanle measurement again yields an effective lifetime of roughly 0.14 ns (open purple circles). While the rotating Hanle geometry does yield a slightly shorter lifetime of 0.09 ns (solid blue squares), this variation is too small (and in the wrong direction) to account for the longer effective lifetime observed for FMR driven spin pumping. We therefore conclude that the enhanced dephasing rate observed in Fig. 4b indicates the FMR-driven and electrically-driven spin injection processes differ.

While the origin of this discrepancy is still an active area of investigation, we note that the observed $\tau_{FMR}$ of 0.6 ns is consistent with previous measurements of the spin lifetime in the bulk silicon channel at these doping level[4,23-25]. Further, the requirement that there be no net charge flow during FMR driven spin injection implies that any forward propagating tunneling process be balanced by an equal and opposite back tunneling process. As shown by the band diagram of spin pumping in Fig. 4a this opens up a potential pathway for coupling *from* the bulk Si states *into* the intermediate states that dominate the three-terminal accumulation voltage[26,28]. The band diagram of electrical spin injection (upper panel of Fig. 4a) shows that this process is strongly suppressed in electrical spin injection due to the finite bias (5 mV in this case) present across the tunnel junction. If we assume for the sake of argument that the spin pumping measurement is in fact sensitive to the bulk spin polarization in silicon, we can calculate the spin current using the relation $\Delta\mu_S = eJ_S \rho \lambda_{sf}$; according to Eq. (1) with $P=0.4$ for Fe and assuming $\eta=0.5$, the spin current density is $J_S \sim 2 \times 10^5$ Am$^{-2}$. This spin current is roughly one order of magnitude smaller than previous reports of spin pumping from conventional ferromagnets into metals.

In summary, we demonstrate spin pumping into a semiconductor through an insulating dielectric. This approach allows observation of precession of the dynamically injected spins and characterization of the effective spin lifetime. Our results directly probe the coherence and phase of the dynamically injected spins and the spin manipulation of that spin ensemble via spin precession, laying the foundation for novel spin pumping based spintronic applications.

**Acknowledgements**

This work is supported by the Center for Emergent Materials at the Ohio State University, a NSF Materials Research Science and Engineering Center (DMR-0820414) (YP, PO, JB, AS, RK, JP and EJH) and by the Department of Energy through grant DE-FG02-03ER46054 (RA and PCH). Technical support is provided by the NanoSystems Laboratory at the Ohio State University. The authors thank Andrew Berger and Steven Tjung for discussions and assistance.

**Reference and notes**

1. Žutić, I., Fabian, J. & Das Sarma, S. Spintronics: Fundamentals and applications. *Rev. Mod. Phys.* **76**, 323 (2004).

2. Awschalom, D. D. & Flatté, M. E. Challenges for semiconductor spintronics. *Nature Phys.* **3**, 153 (2007).

3. Appelbaum, I., Huang, B., & Monsma, D. J. Electronic measurement and control of spin transport in silicon. *Nature* **447**, 295 (2007).

4. Dash, S. P., Sharma, S., Patel, R. S., de Jong, M. P. & Jansen, R. Electrical creation of spin polarization in silicon at room temperature. *Nature* **462**, 491 (2009).

5. Jedema, F. J., Heersche, H. B., Filip, A. T., Baselmans, J. J. A. & van Wees, B. J. Electrical detection of spin precession in a metallic mesoscopic spin valve. *Nature* **416**, 713 (2002).

6. Lou, X. *et al.* Electrical detection of spin transport in lateral ferromagnet semiconductor devices. *Nature Phys.* **3**, 197 (2007).

7. Jonker, B. T., Kioscoglou, G., Hanbicki, A. T., Li, C. H., & Thompson, P. E. Electrical spin-injection into silicon from a ferromagnetic metal/tunnel barrier contact. *Nature Phys.* **3**, 542 (2007).

8. Kato, Y. K., Myers, R. C., Gossard, A. C. & Awschalom, D. D. Observation of the Spin Hall Effect in Semiconductors. *Science* **306**, 1910 (2004).

9. Valenzuela, S. O. & Tinkham, M. Direct electronic measurement of the spin Hall effect. *Nature* **442**, 176 (2006).

10. Liu, L. *et al.* Spin-Torque Switching with the Giant Spin Hall Effect of Tantalum. *Science* **336**, 555 (2012).

11. Tserkovnyak Y., Brataas, A. & Bauer, G. E. W. Enhanced Gilbert damping in thin ferromagnetic Films. *Phys. Rev. Lett*. **88**, 117601 (2002).

12. Saitoh, E., Ueda, M., Miyajima, H. & Tatara, G. Conversion of spin current into charge current at room temperature: Inverse spin-Hall effect. *Appl. Phys. Lett.* **88**, 182509 (2006).

13. Tserkovnyak Y., Brataas, A. & Bauer, G. E.W. & Halperin, B. I. Nonlocal magnetization dynamics in ferromagnetic heterostructures. *Rev. Mod. Phys.* **77**, 1375 (2005).

14. Kajiwara, Y. *et al*. Transmission of electrical signals by spin-wave interconversion in a magnetic insulator. *Nature* **464**, 262 (2010).

15. Kurebayashi, H. *et al*. Controlled enhancement of spin-current emission by three-magnon splitting. *Nature Materials* **10**, 660 (2011).

16. Sandweg, C. W. *et al*. Spin Pumping by Parametrically Excited Exchange Magnons. *Phys. Rev. Lett.* **106**, 216601 (2011).

17. Czeschka, F. D. *et al*. Scaling behavior of the spin pumping effect in ferromagnet-platinum bilayers. *Phys. Rev. Lett.* **107**, 046601 (2011).

18. Ando, K. *et al.* Inverse spin-Hall effect induced by spin pumping in metallic system. *J. Appl. Phys.* **109**, 103913 (2011).

19. Ando, K. *et al*. Electrically tunable spin injector free from the impedance mismatch problem. *Nature Materials* **10**, 655 (2011).

20. Ando, K. & Saitoh, E. Observation of the inverse spin Hall effect in silicon. *Nature Comm.* **3**, 629 (2012).

21. Costache, M. V. *et al.* Electrical detection of spin pumping due to the precessing magnetization of a single ferromagnet. *Phys. Rev. Lett.* **97**, 216603 (2006).

22. Heinrich, B. *et al*. Spin pumping at the magnetic insulator (YIG)/normal metal (Au) interfaces. *Phys. Rev. Lett.* **107**, 066604 (2011).

23. Dash, S. P. *et al*. Spin precession and inverted Hanle effect in a semiconductor near a finite-roughness ferromagnetic interface. *Phys. Rev. B* **84**, 054410 (2011).

24. Li, C. H., van't Erve, O. & Jonker, B. T. Electrical injection and detection of spin accumulation in silicon at 500K with magnetic metal/silicon dioxide contacts. *Nature Comm.* **2**, 245 (2011).

25. Gray, N. W. & Tiwaria, A. Room temperature electrical injection and detection of spin polarized carriers in silicon using MgO tunnel barrier. *Appl. Phys. Lett.* **98**, 102112 (2011).

26. Pu, Y. *et al*. Correlation of electrical spin injection and non-linear charge-transport in Fe/MgO/Si. *Appl. Phys. Lett.* **103**, 012402 (2013).

27. Jeon, K. *et al*. Electrical investigation of the oblique Hanle effect in ferromagnet / oxide / semiconductor contacts. *arXiv* 1211.3486 (2013).

28. Tran, M. *et al*. Enhancement of the spin accumulation at the interface between a spin-polarized tunnel junction and a semiconductor. *Phys. Rev. Lett.* **102**, 036601 (2009).

29. Sasaki, T., Oikawa, T., Suzuki, T., Shiraishi, M., Suzuki, Y. & Noguchi, K. Comparison of spin signals in silicon between nonlocal four-terminal and three-terminal methods. *Appl. Phys. Lett.* **98**, 012508 (2011).

30. As discussed in Ref. 26, Eq. (1) is derived assuming a linear tunneling model, and might underestimate the actual value of $\Delta\mu_S$ at higher bias if the current depends super-linearly on applied bias.

# Figure legends

**Figure 1 | Experimental setup**

(**a**) Schematic of experimental setup. (**b**) Diagram of spin accumulation and spin-resolved voltage $V_S$. (**c**) FMR intensity (upper panel; arrows indicate state of the Fe magnetization) and spin-resolved voltage $V_S$ (lower panel) as a function of $H_{DC}$.

**Figure 2 | RF power- and magnetic field- dependence**

(**a**) FMR intensity (upper panel) and spin-resolved voltage (lower panel) as a function of RF power; (**b**) Solid symbols: $V_S$ vs. $H_{DC} - H_{FMR}$ when $H_{DC}$ is parallel or anti-parallel with the *x*-axis; open symbols indicate the voltage between two Au/Si reference contacts; all data is measured under the same experimental conditions. A background offset of ~2 µV has been subtracted from all data.

**Figure 3 | Experiments with increasing H$_z$**

(**a**) FMR intensity spectra at various magnetic field orientations $\alpha$ as described in the text; (**b**) $V_S$ vs. $H_{DC}$ measured at the same set of magnetic field orientations. The shift in FMR center frequency tracks the expected magnetization anisotropy of the Fe thin film, see text.

**Figure 4 | Hanle effect measurements**

(**a**) Schematics of experimental setup and band diagram for three-terminal electrical spin injection (upper panel) and spin pumping (lower panel); (**b**) Hanle effect as a function of $H_z$ for three-terminal (open circles) and spin pumping (solid circles), solid lines are Lorentzian fits yielding 0.11 ns and 0.6 ns, respectively; solid triangles are FMR absorption as a function of $H_z$, the red dashed line is a guide to the eye and the scale bar represents 10% variation; Black dashed lines are simulated using Eq. (4) with $\tau = \infty$, $H_x$ = 248G (dash dot) and 310G (dash), respectively, see Supplementary Information. (**c**) Hanle effect measured in a control sample by the three-terminal method, open circles are measured with magnetic field applied out of plane, solid squares are obtained using same magnetic field configuration as for FMR driven spin pumping; lines are Lorentzian fits yielding 0.14 ns and 0.09 ns, respectively.

**Figures**

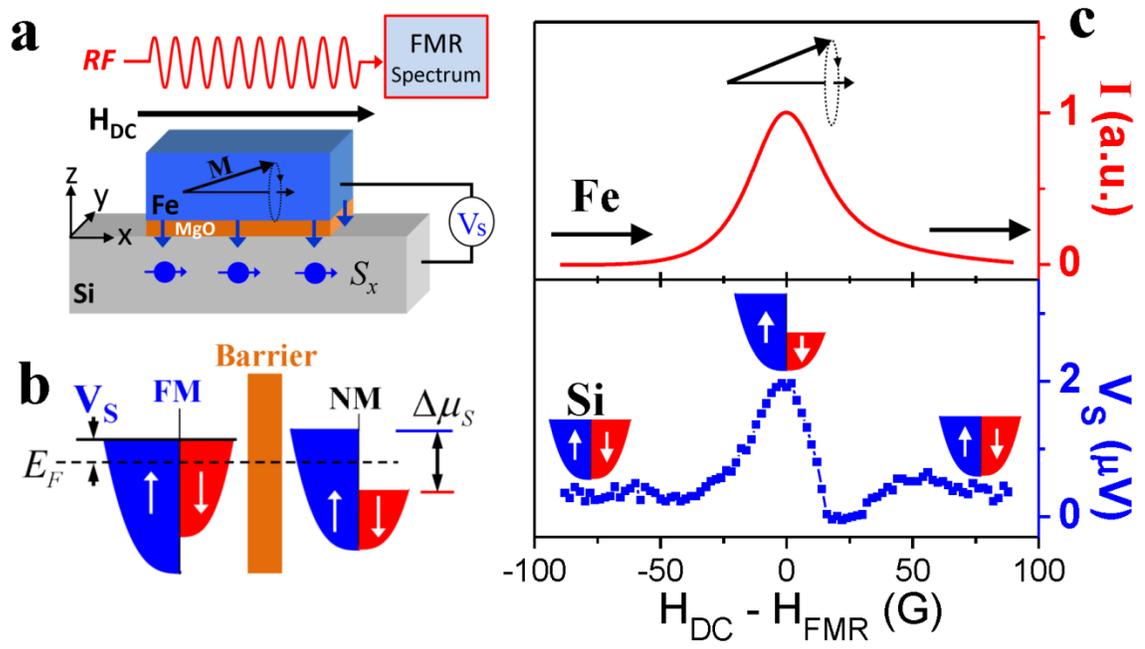

**Figure 1** Y. Pu *et al.*

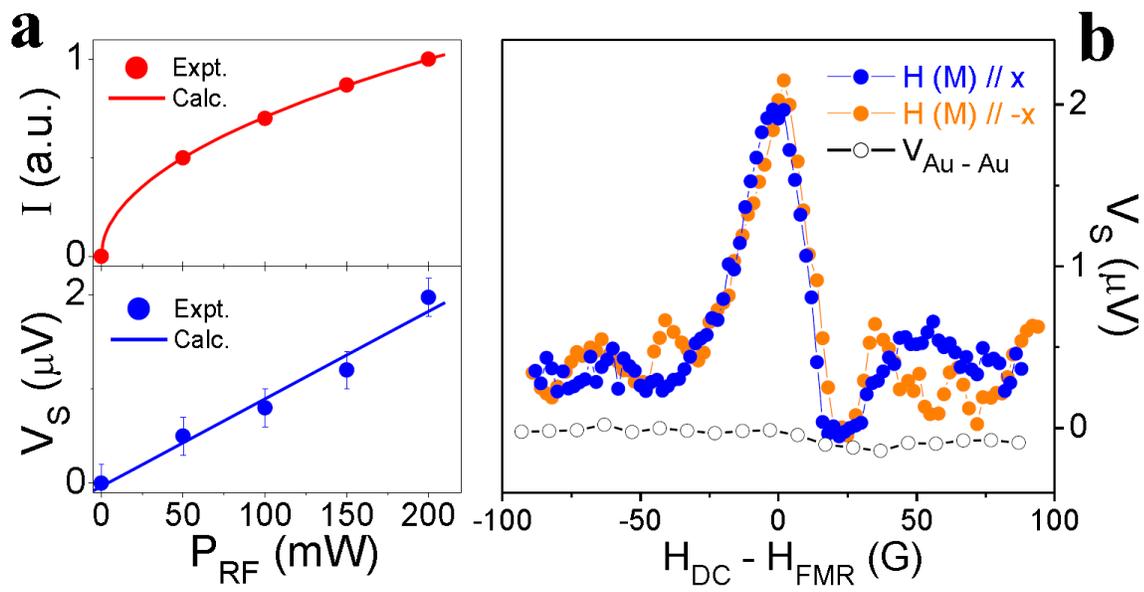

**Figure 2** Y. Pu *et al.*

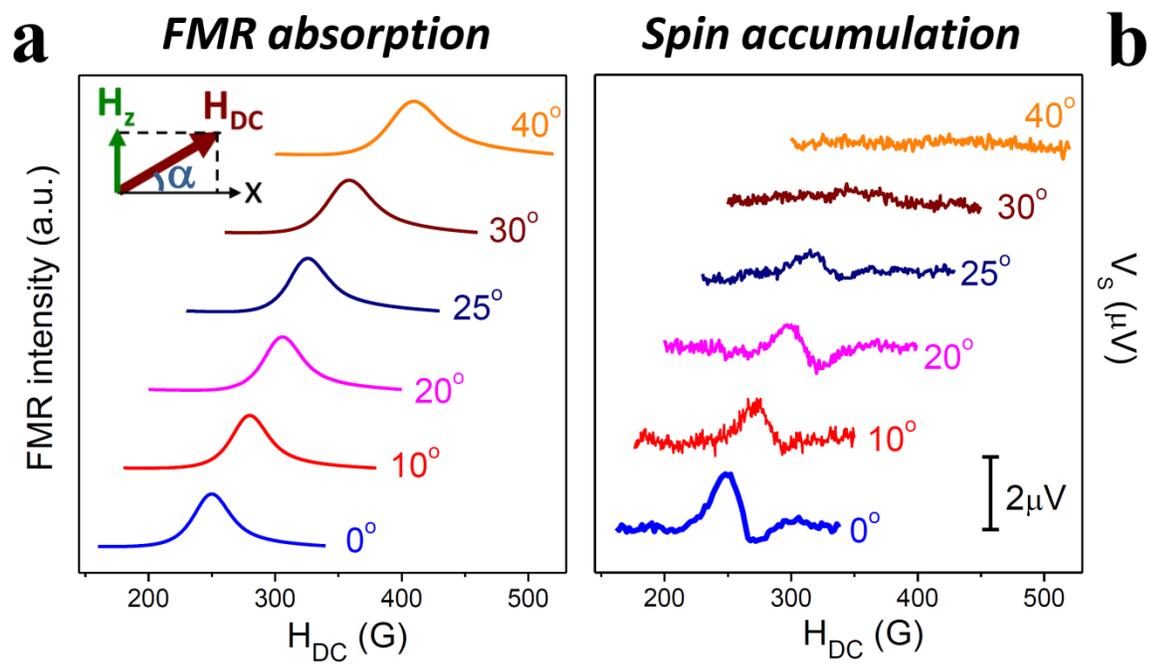

**Figure 3** Y. Pu *et al.*

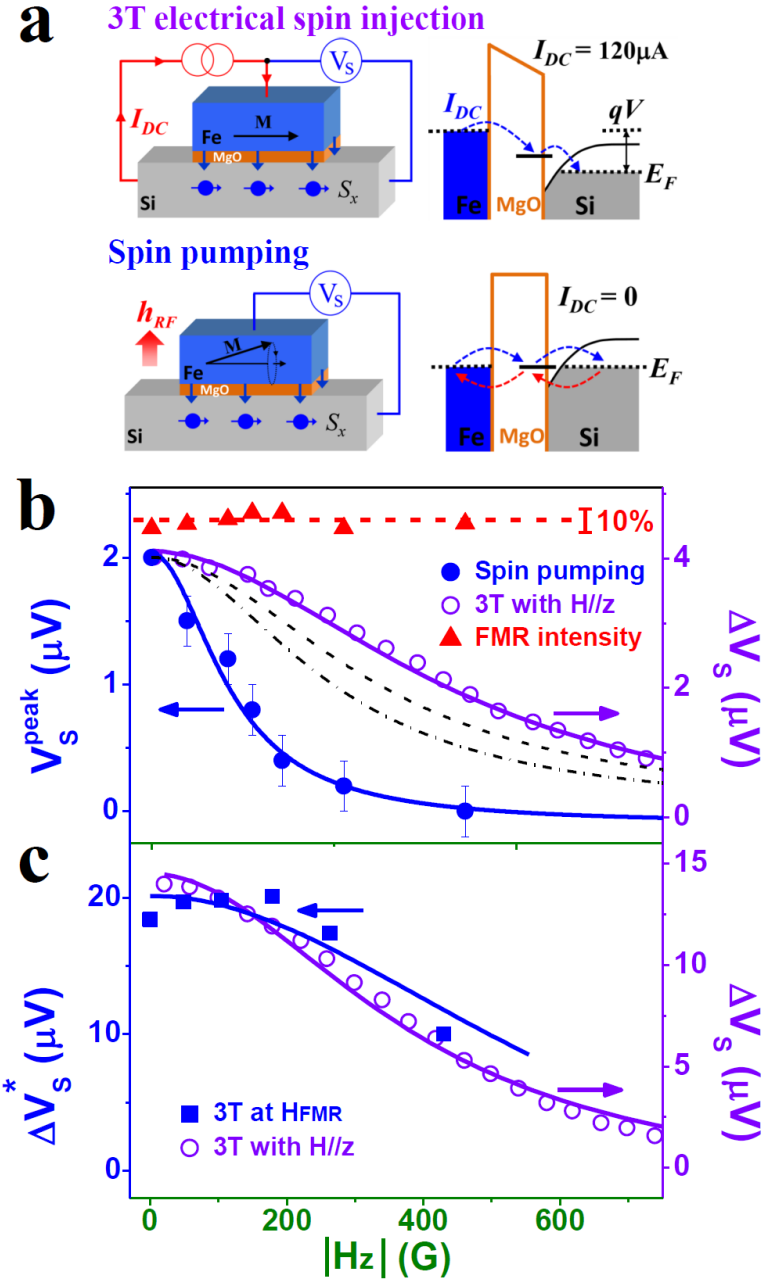

**Figure 4** Y. Pu *et al.*

*Supplementary Information*

# Spin accumulation detection of FMR driven spin pumping in silicon-based metal-oxide-semiconductor heterostructures

Y. Pu[1], P. M. Odenthal[2], R. Adur[1], J. Beardsley[1], A. G. Swartz[2], D. V. Pelekhov[1], R. K. Kawakami[2], J. Pelz[1], P. C. Hammel[1], E. Johnston-Halperin[1]

[1]Department of Physics, The Ohio State University, Columbus, Ohio 43210

[2]Department of Physics and Astronomy, University of California, Riverside, California 92521

### A. Linear I-V of Fe/MgO/Si contact at room temperature

The I-V characteristic of the Fe/MgO/p-Si contact is linear at room temperature, as shown in Fig. S1, indicating that in this regime the contact resistance is dominated by the MgO insulating barrier and contribution from the Schottky barrier is negligible. The linear fit gives 40.9Ω resistance with about 25mΩ uncertainty.

### B. Impact of interface roughness and in-plane external magnetic field

In contrast to traditional three-terminal spin accumulation measurements, for the FMR driven measurements described in Figs. 3 and 4 it is necessary to apply an external magnetic field *both* parallel and perpendicular to the magnetization. As described in the main text, the parallel component of the field is necessary to satisfy the conditions of magnetic resonance and the perpendicular component contributes to the decay of spin accumulation, allowing for a Hanle-style measurement of the effective spin lifetime. However, in consulting the literature[23,27] it becomes evident that this geometry potentially raises an additional concern regarding the interpretation of this data. Specifically, the in plane component of the magnetic field will itself induce an "inverted-Hanle" effect wherein the measured spin accumulation *rises* with the magnitude of the parallel component of the magnetic field (Fig. S2a). The accepted interpretation of this effect is an annealing of fluctuations in the magnetization induced by surface roughness of the magnetic layer in large external fields[23,27].

We measure the spin accumulation on a control sample via 3T electrical spin injection, as shown in Fig. S2a, where magnetic field is applied in xz-plane with orientation ranging from in-plane (*H//x*, 0 degree) to out of plane (*H//z*, 90 degree). Using the 3T data of $\Delta V_S(H = H_{FMR})$, where $H_{FMR}$ for given magnetic field orientation is obtained by spin pumping experiment as shown in Figure 3, we can first directly compare electrical spin injection and spin pumping under the *same* experimental configuration, as discussed in main text.

To get a quantitative understanding of the angular dependence shown in Fig. S2a, one can start with a general formula:

$$\frac{\partial \mathbf{S}}{\partial t} = \mathbf{S} \times \boldsymbol{\omega} + D\nabla \mathbf{S} - \frac{\mathbf{S}}{\tau} \quad (S1)$$

where $\omega$ is the Larmor frequency, $D$ is the spin diffusion constant and $\tau$ is the spin lifetime. In the 3T geometry the impact of spin diffusion is usually believed to be negligible[23,27], from this one can obtain an analytical solution under arbitrary applied magnetic field:

$$S_x = S_0 \left\{ \frac{\omega_x^2}{\omega_{total}^2} + \frac{\omega_y^2 + \omega_z^2}{\omega_{total}^2} \frac{1}{1+(\omega_{total}\tau)^2} \right\} \quad (S2)$$

where $\omega_i = H_i g_{eff} \mu_B / \hbar$ $(i=x,y,z)$, representing each component of the applied magnetic field. Figure S2b shows the simulation according to Equation (S2) with $\tau = 0.14$ns as obtained by Lorentzian fit. Clearly the model fails to explain the data shown in Fig. S2a, especially the observation that at certain orientations the measured spin accumulation *rises* with the applied magnetic field increasing.

As pointed out by previous studies[23,27] stray magnetic fields from the injector due to interface roughness should strongly impact on the Hanle-style measurements. The total magnetic field should be taken as $H_i = H_i^{ext} + H_i^{ms}$ $(i=x,y,z)$, which represents the contribution from external and magnetic-stray fields. The stray field strength is taken to have a spatial variation $H_i^{ms}(x) = H_i^{ms}(0) \cdot \cos(2\pi x/\lambda)$, where $\lambda$ is the typical length scale (~20nm) of the surface roughness[23,27]. Assuming the spin diffusion length is much longer than $\lambda$ we average the total magnetic field over a full period of $\lambda$, i.e. $\overline{H}_i^2 = (H_i^{ext})^2 + (\overline{H}_i^{ms})^2$, we therefore have a formula:

$$S_x = S_0 \left\{ \frac{(\overline{H}_x^{ms})^2 + (H_x^{ext})^2}{H_{total}^2} + \frac{(\overline{H}_\perp^{ms})^2 + (H_z^{ext})^2}{H_{total}^2} \frac{1}{1+(g\mu_B/\hbar)^2 H_{total}^2 \tau^2} \right\} \quad (S3)$$

where $\overline{H}_x^{ms}$ and $\overline{H}_\perp^{ms}$ represent the averaged stray field parallel or perpendicular to the injected spins, respectively. Figure S2c shows the simulation according to Equation (S3) with parameters $\tau = 0.9$ns, $\overline{H}_x^{ms} = 270$G and $\overline{H}_\perp^{ms} = 440$G. The simulation qualitatively agrees with the experiment, but shows some systematic deviations especially in the low-field regime.

Although the Equation (S1) is generally accepted, and in principle rigorous analysis can be done with spin precession, spin diffusion and spin flip involved, a well-established approach to determine the intrinsic spin lifetime using the local spin detection geometry is still lacking. A precise determination of the intrinsic spin lifetime in our sample is beyond the scope of this report; we treat the lifetime obtained by the simple Lorentzian fit as an effective spin lifetime or spin dephasing time, which represents the decay rate of average spin accumulation under applied perpendicular magnetic field.

## C. Simulation on the Hanle effect under FMR condition

As indicated by the FMR spectrum, at FMR there are varying $x$- and $z$- components of the applied magnetic field with different field orientation $\alpha$. The table below is a summary:

| $\alpha$ (deg.) | 0 | 10 | 20 | 25 | 30 | 40 | 60 |
|---|---|---|---|---|---|---|---|
| $H_x$ (G) | 250 | 276 | 287 | 295 | 310 | 302 | 248 |
| $H_z$ (G) | 0 | 49 | 105 | 138 | 179 | 253 | 430 |

As shown in the table, $H_z$ increases monotonically with $\alpha$ and $H_x$ is in range of 248 – 310 G (roughly constant to maintain the conditions for magnetic resonance), both should impact on the Hanle effect. The black dashed curves shown in Fig. 4(b) are simulated using Eq. (4) with $\tau = \infty$ and $H_x$ = 248G, 310G respectively. In the situation that Eq. (4) is valid, finite values of the spin lifetime should give a weaker $H_z$-dependence than the simulation curve.

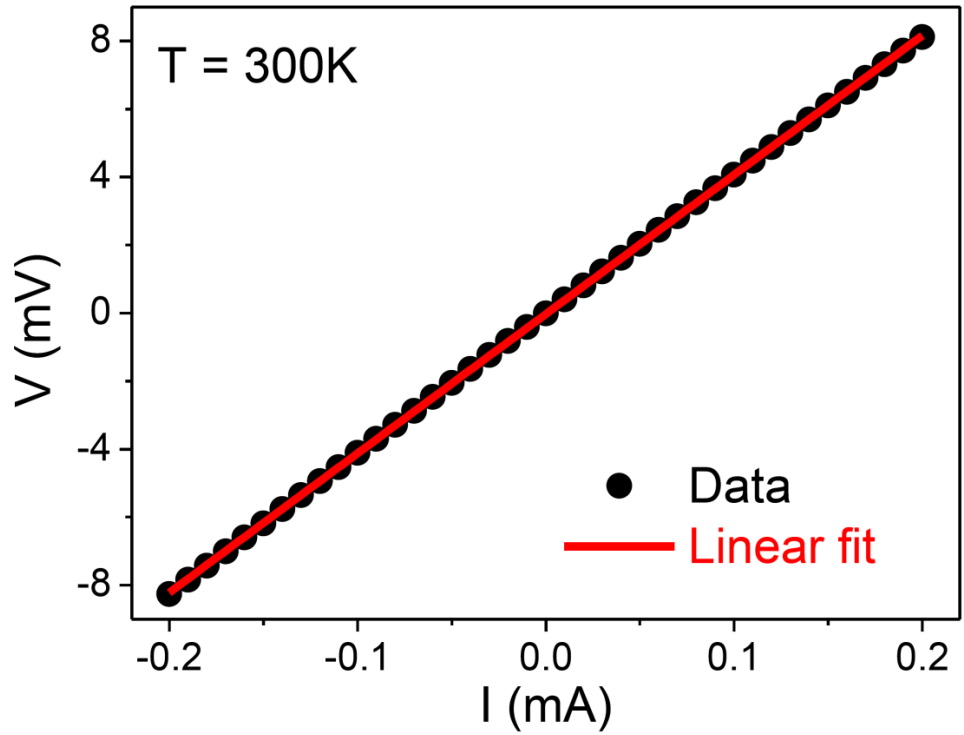

**Figure S1:** Plot of current vs. voltage of the Fe/MgO/p-Si contact at room temperature, symbols are data and the solid line is a linear fit.

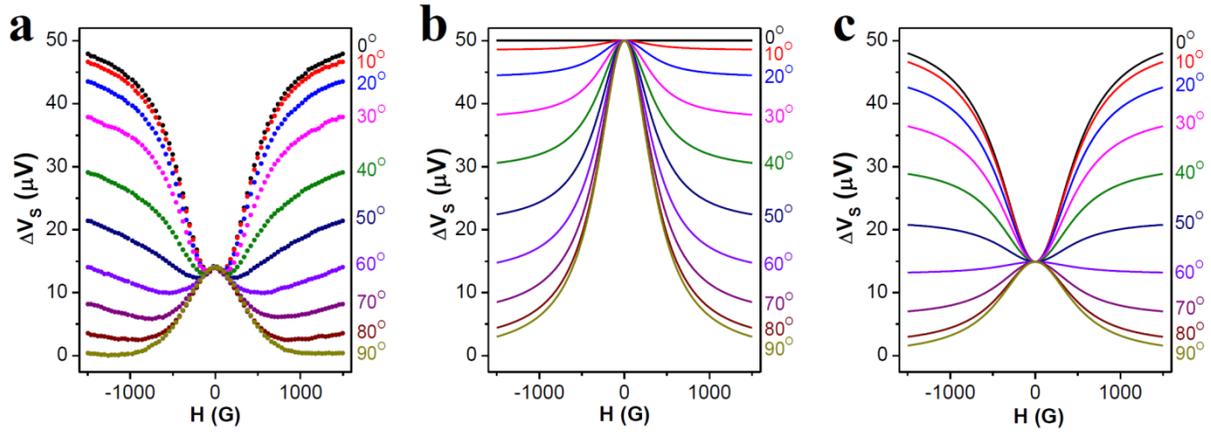

**Figure S2:** (**a**) Spin accumulation from 3T electrical spin injection, magnetic field is applied in different orientations, ranging from in-plane (*H//x*, 0 degree, top curve) to out of plane (*H//z*, 90 degree, bottom curve); (**b**) Simulations according to Equation (S2) with $\tau = 0.14$ ns, assuming no stray field; (**c**) Simulations using Equation (S3) with parameters $\tau = 0.9$ ns, $\bar{H}_x^{ms} = 270$ G and $\bar{H}_\perp^{ms} = 440$ G.